\documentclass[12pt,epsfig,multirow]{article}
\usepackage{graphicx}
\usepackage[usenames]{color}
\newcommand{\be}{\begin{eqnarray}}
\newcommand{\ee}{\end{eqnarray}}

\def\dis{\displaystyle}
\def\barr{\begin{array}}
\def\earr{\end{array}}

\begin{document}

\thispagestyle{empty}
\begin{flushright}

\end{flushright}
\bigskip

\begin{center}
{\Large \bf Isospin nonconservation for fp-shell nuclei by spectral distribution theory}

\vspace{.5in}

{\bf {Kamales Kar$^{\star, a}$ and Sukhendusekhar Sarkar$^{\dagger,b}$}}
\vskip .5cm
$^\star${\normalsize \it Ramakrishna Mission Vivekananda University, }
{\normalsize \it Belur Math, Howrah 711202, India}\\
\vspace{0.05in}
$^\dagger${\normalsize \it Indian Institute of Engineering Science and Technology, Shibpur, Howrah 711103, India}\\
\vskip 0.4cm

\vskip 1cm

{\bf ABSTRACT}
\end{center}
The one- plus two-body isospin nonconserving nuclear interactions
are included in the prediction of ground state energies of fp shell
nuclei using spectral distribution theory. This in turn is used
to calculate the linear term in the isobaric mass-multiplet
equation and the predictions are then compared to experimental values
after the addition of the Coulomb contribution.
The agreement is found to be reasonable as observed for sd shell
nuclei earlier. One also sees that in this method the contribution
to the linear term comes almost completely from the one body isovector
Hamiltonian and that results in a huge simplification of the problem.

\vfill

\noindent $^a$ email: kamales.kar@gmail.com

\noindent $^b$ email: ss@physics.iiests.ac.in ; sukhendusekhar.sarkar@gmail.com

\newpage
Though isospin symmetry is one of the most well-known symmetries in nuclear
structure and using the goodness of isospin quantum number considerable
simplifications are obtained in nuclear structure calculations, the 
small breaking of this symmetry by nuclear interactions has been observed
and experimentally measued over the years. Wigner \cite{Wigner-57} was the
first one to postulate an isobaric mass multiplet equation (IMME) which
quantitaively describes the isospin non-conservation and is given by

\be
M(\alpha,T,T_z)=a(\alpha,T)+ b(\alpha,T) T_z+ c(\alpha,T) T_z^2
\ee
 
where $M(\alpha,T,T_z)$ stands for the masses of the nuclei in a multiplet
with fixed isospin $T$ and $T_{z} (=(N-Z)/2)$ takes values -$T$, -$T$+1,...., $T$-1, $T$.
Thus the equation stands for the masses of (2T+1) isobars.
The parameters `a' , `b' and `c' are often evaluated by fits to experimental
values.
For microscopic nuclear structure calculations like the shell model one 
writes the total Hamiltonian ($H$) as not only the symmetry-preserving
isoscalar one- and two-body parts ($H^{(0)}$) but with the addtion of the one- 
and two-body isovector ($H^{(1)}$) and the two-body isotensor ($H^{(2)}$) parts.
Many shell model and other calculations have been carried out for
light nuclei in the $p$-shell, $sd$-shell, $fp$-shell \cite{Ormand-89}
\cite{Lam-13} and the theroetical predictions for `b' and `c' of IMME compared
to the experimental numbers. Ormand and Brown have pointed out that the
the parameter `b' times $`T_z'$ is exactly equal to the contribution to the ground state
coming from the isovector Hamiltonian \cite{Ormand-89}. 

Spectral distribution theory 
 \cite{French-66} \cite{French-67} \cite{FR-71}
\cite{CFT-71} \cite{Kotahq-10}
\cite{Brody-81} \cite{French-82} \cite{Kota-89}
describes the statistically averaged nuclear structural properties avoiding
explicit diagonalisation of the Hamiltonian and calculates the relevant
quantities evaluating traces of operators and product of operators in the 
shell model spaces. It has been applied successfully to calculate spectra
and energy-averaged transition strengths of light nuclei \cite{Kota-89}
\cite{Gomez-11}.
This theory has recently been extended to include non-isoscalar one- and
two-body interactions and a methodology developed to calculate the linear
 term in the IMME \cite{Kar-15}. Examples of nuclei in the $sd$-shell 
show that the theory works quite well in predicting the parameter `b'
in IMME. In this work we consider nuclei belonging to the $fp$-shell and
observe that the theory is reasonably successful for these nuclei as well.
We also see that using the Ormand-Brown interaction and their parametrisation \cite{Ormand-89} the prediction for the linear term changes very
little when one neglects the two body parts in the isovector and 
isotensor Hamiltonians. This was observed earlier for $sd$-shell nuclei
too \cite{Kar-15}. 

In spectral distribution theory one observes that with `m' valence
particles coupling to total isospin `T', the eigenvalue density
 of all realistic Hamiltonians in the configuration shell model spaces 
is very close to Gaussian as long as the dimension of each
of this space is large. So the energy intensities $I_{{\bf m},T}(E)$, 
defined as the energy eigenvalue density $\rho_{{\bf m},T}$ times the dimension
of the configuration space $d(\bf m,T)$ adding up to give the total
intensity $I_{m,T}(E)$, are functions of only two quantities,
 the centroid and the
width for each configuration. So the total intensity is given by

\be
I_{m, T}(E)= \dis \sum_{\bf m} I_{{\bf m},T} (E)= \dis \sum_{\bf m} d({\bf m},T) \rho_{{\bf m},T} (E)
\ee

Once one is able to calculate the fixed T configuration centroids and widths in this theory
by using the unitary group structure of the shell model spaces with
`m' particles distributed over 'l' valence orbits (for the 
$fp$-shell case the orbits are $0f_{5/2}$, $0f_{7/2}$, $1p_{3/2}$ 
and $1p_{1/2}$) one knows the 
eigenvalue densities and the intensities. 

Though the spectral distribution theory describes
the global averaged properties, it can give information of the 
microscopic states like the ground state and the low-lying spectra. The ground 
state energy of a nucleus with a fixed number of valence particles and 
isospin  is evaluated by a method named the Ratcliff procedure 
\cite{Ratcliff-71}. If $d_i$ is the degeneracy of the $(i-1)$-th 
excited state ($d_1$ being the degeneracy of the ground state) the
energy of that state $\bar{E_i}$ is obtained by inverting the equation

\be
\dis \sum_{\bf{m}} \int_{-\infty}^{\bar{E_{i}}} I_{{\bf m},T}(E) dE = \dis \sum_{k=1}^{i-1} d_k + d_{i}/2
\ee

The calculated energy of the ground state $\bar{E_1}$ will be denoted 
by the more familiar symbol
$E_{g}$. We calculate the ground state energies by spectral ditributions
 for the cases $(m=3,T=1/2)$, $(m=5,T=1/2)$ and $(m=6,T=1)$ with $^{40}Ca$
as the closed core, using the $fp$-shell FPV interaction \cite{vanHees-81}.
These calculations are constrained to have only one particle outside the 
$0f_{7/2}$ orbit following Table 1 of Ref. \cite{Ormand-89}  and so we choose only such configurations for our
calculations for all the three cases. We first convert the two-body matrix
elements from the (JT) form in the 4 orbits to the proton-neutron form in
8 orbits (4 proton orbits plus 4 neutron orbits). The expressions for the pp,
nn and pn/np matrix elements for the isoscalar interactions are 
wellknown \cite{Mugambi-70} and those
for the isovector and isotensor interaction are given by Ormand and 
Brown \cite{Ormand-89} and discussed by Kar and Sarkar \cite{Kar-15}.
The centroids and widths in the configurations are obtained by 
evaluating traces using the group theoretical structure of the configuration-pn
spaces first \cite{Kota-89} and then in configurations with fixed isospin
by the method of subtraction of the traces \cite{Mugambi-70} \cite{Sarkar-92}
 \cite{Kar-15}. 

Table 1 gives the centroids and widths in all the 4 fixed
isospin configurations considered with the FPV interaction for the isoscalar
Hamiltonian and compares them with the values for the total Hamiltonian.
 We observe, that in all the cases, with the addition
of the isovector and isotensor parts, the centroids move away
from the values with only isoscalar but the widths remain almost 
the same.
 
In Table 2 we present the ground state energies of the 3, 5 and 6
particle cases calculated by the method described above. Values are 
given for both the isoscalar $H$ and the total $H$ in fixed isospin
spaces as well as for spaces with fixed proton and neutron numbers
(taking $T_z$=$T$). The coefficient `b' in the linear term in the
IMME is obtained by equating the difference between the calculated
lowest $(m,T,J)$ state energy with the isoscalar plus isovector Hamiltonian
($H^{(0)}$+$H^{(1)}$) and only the isoscalar Hamiltonian ($H^{(0)}$) to
$b T_z$. The contribution of the Coulomb term to `b' can be evaluated
by using the equation (25) of Lam, Smirnova and Caurier \cite{Lam-13}.
However in this work we use for `b' coming from
Coulomb energy, the expression given by the equation (10) in
the work of Ormand \cite{Ormand-97} which is a global fit with good
accuracy. The experimental values are as quoted in Ormand and Brown
\cite{Ormand-89}. For the 5 particle case alongwith the $J=7/2$ ground 
state the table also presents
the values for the first and second excited states with $J=3/2$ and
$J=5/2$ respectively. We find the agreement of our results with experimental 
values reasonable keeping in mind that spectral distribution theory
is constructed to describe the statistically averaged global 
properties of nuclei.


\begin{table}
\begin{tabular}{|c|c|c|c|c|c|}
\hline
Fixed T  & Dimension & Isoscalar H & Isoscalar H & Total H  & Total H     \\
configuration  &   & centroid (MeV) & width (MeV) & centroid (MeV)& width (MeV) \\
\hline
\hline
 (1200) & 384 &  -21.12  &  2.16   &  -24.83  &  2.16   \\
 (0300) & 168 &  -27.39  &  2.07   &  -31.10  &  2.07   \\
 (0210) & 256 &  -25.34  &  1.85   &  -29.00  &  1.85   \\
 (0201) & 128 &  -23.55  &  1.88   &  -27.22  &  1.88   \\
\hline
 (1400) & 4284 &  -42.42 &  3.37    & -46.13  &  3.37   \\
 (0500) & 1080 &  -48.46 &  3.27    & -52.17  &  3.27   \\
 (0410) & 2856 &  -46.05 &  3.13    & -49.73  &  3.13   \\
 (0401) & 1428 &  -44.37 &  3.15    & -48.05  &  3.15   \\
\hline
 (1500) & 9072 &  -52.81  &  3.71   &  -60.22  &  3.71   \\
 (0600) & 1512 &  -59.11  &  3.49   &  -66.52  &  3.49   \\
 (0510) &  6048 &  -56.34  &  3.48   &  -63.70  &  3.48   \\
 (0501) &  3024 &  -54.69  &  3.50   &  -62.05  &  3.50   \\
\hline
 \end{tabular}
\caption{
Centroids and widths of the isoscalar Hamiltonian compared to the
centroids and widths of the total Hamiltonian (including the isovector
and istensor parts) in the fixed-T configurations with 3, 5 and 6  particles
in $0f-1p$ shell with $T=1/2$, $T=1/2$ and $T=1$ respectively.
The notation $(m_1,m_2,m_3,m_4)$ stands for the configuration with $m_1$
particles in orbit $0f_{5/2}$, $m_2$  particles in orbit $0f_{7/2}$, $m_3$
particles in orbit $1p_{3/2}$ and $m_4$ in orbit $1p_{1/2}$. 
}
\end{table}

\begin{table}
\begin{tabular}{|c|c|c|c|c|c|c|c|}
\hline
(m,T,J) & LSE for & LSE for  & LSE for& $b$ from  & $b$ for & Total $|b|$ & Observed    \\
       & $H^{(0)}$&Total H & $H^{(0)}+H^{(1)}$ & Total H  &Coulomb  & (MeV)& $|b|$      \\
      &(MeV) &(MeV) & (MeV)   &(MeV)  &(MeV) & &(MeV)      \\
\hline
\hline
(3,1/2,7/2) &  -31.52 (-31.30) &   -35.21 (-35.00)  &-35.21 &  -7.38   & -0.41   &    7.79    & 7.650 \\
(5,1/2,7/2) & -57.30 (-56.97)  &  -61.01 (-60.68) &-61.00  & -7.40   & -0.68   &    8.08    & 7.914 \\
(5,1/2,3/2)$^{*}$ & -56.56 (-56.28)      &   -60.26 (-59.98)   & -60.26    & -7.40  &  -0.68   &  8.08   & 7.934  \\
(5,1/2,5/2)$^{*}$ & -55.92 (-55.69)      &  -59.63 (-59.39)     & -59.62   & -7.40  &  -0.68   &  8.08   & 7.930  \\
(6,1,0) & -71.16 (-70.90)   &   -78.57 (-78.31) & -78.57 &  -7.41   & -0.81   &    8.22    & 8.109  \\
\hline
\end{tabular}
\caption{
The parameter $`b'$ of IMME coming from nuclear interactions from evaluation
of the lowest state energies (LSE) calculated by
spectral distributions. The Coulomb contribution is calculated using eq (25)
of ref \cite{Lam-13}. The numbers in the parentheses
are LSEs calculated in the $(m_p,m_n)$ spaces.
The states with the
$^{(*)}$ are not the ground states but the lowest $(m=5,T=1/2)$ states
with $J=3/2$ and $J=5/2$ whereas the other 3 are ground states
}

\end{table}

\begin{table}
\begin{tabular}{|c|c|c|c|c|}
\hline
A  &   T   &  $|b|$ by Spectral Distributions & $|b|$ by Spectral Distributions & Observed $|b|$  \\
   &       &                  & with Isoscalar and 1-body Isovector &     \\
\hline
  43  &  1/2   &   7.79   &   7.79     &   7.65     \\
  45  &  1/2   &   8.08   &   8.09     &   7.91     \\
  46  &   1    &   8.22   &   8.23     &   8.11     \\
\hline
\end{tabular}
\caption{
Comparison of $|b|$ using isoscalar with only one-body isovector Hamiltonian with
$|b|$ from the full H 
}
\end{table}


Finally in Table 3, we show the results for `b' when we include
only the one-body isovector $H^{(1)}$ and not the two-body part in
our calculations. We compare them with the results given in Table 2
where both the one- and two-body parts of $H^{(1)}$ are included.
One observes that there is hardly any change in the results when
one drops the two-body isovector part. This feature was observed
for the $sd$-shell nuclei in the earlier work of Kar and Sarkar
\cite{Kar-15}. This originates from the fact the overall multiplicative
coefficient in the two-body isovector Hamiltonian of  
Ormand and Brown \cite{Ormand-89} which is used by us, is
very small. But this parametrisation of the Hamiltonian had resulted
in shell model predictions which were very accurate \cite{Ormand-89}.
The fact that the two body part need not be included
for calculating the linear term of IMME is an enormous simplification
and may be followed up for other nuclei in future.

In conclusion, we observe that the linear term in the isobaric 
mass-multiplet equation for isospin-breaking can be calculted by the spectral
distribution theory and its predictions agree well with the available
data for the $fp$-shell nuclei as well, seen earlier to work well
for the $sd$-shell nuclei.


\begin{thebibliography}{99}

%
\bibitem{Wigner-57} E.P. Wigner in Proceedings of the Robert A. Welch
Foundation Conference in Chemical Research, Vol 1, edited by W.O. Milligan
(Welch Foundation, Houston, 1957) p 86

\bibitem{Ormand-89} W.E. Ormand and B.A. Brown, Nucl. Phys. A {\bf 491} 1
(1989)

\bibitem{Lam-13} Yi Hua Lam, N.A. Smirnova and E. Caurier, Phys. Rev. C {\bf 87}
 054304 (2013)

\bibitem{French-66} J.B. French, Phys. Lett. {\bf 23} 248 (1966)

\bibitem{French-67} J.B. French, Phys. Lett. {\bf 26} 75 (1967)

\bibitem{FR-71} J.B. French and K. Ratcliff, Phys. Rev. C {\bf 3} 94 (1971)

\bibitem{CFT-71} F.S. Chang, J.B. French and T.H. Thio, Ann. Phys. (NY)
{\bf 66} 137 (1971)

\bibitem{Kotahq-10} V.K.B. Kota and R.U. Haq, "Spectral Distributions in 
Nuclei and Statistical Spectroscopy", World Scientific Publishing Co., 2010,
and the collection of papers therein

\bibitem{Brody-81}
T.A. Brody, J. Flores, J.B. French, P.A. Mello, A. Pandey and S.S.M. Wong,
Rev. Mod. Phys. {\bf 53} 385 (1981)

\bibitem{French-82}
J.B. French and V.K.B. Kota, Ann. Rev. Nucl. Part. Sci. {\bf 32} 35 (1982)

\bibitem{Kota-89}
V.K.B. Kota and K. Kar, Pramana- J. Phys. {\bf 32} 647 (1989)




%



\bibitem{Gomez-11}
J.M.G. Gomez, K. Kar, V.K.B. Kota, R.A. Molina, A. Relano and J. Retamosa,
Phys. Rep. {\bf 499} 103 (2011)

\bibitem{Kar-15}
K. Kar and S. Sarkar, J. Phys. G {\bf 42} 055110 (2015)

\bibitem{Ratcliff-71} K.F. Ratcliff, Phys. Rev. C {\bf 3} 117 (1971)

\bibitem{vanHees-81} A.G.M. van Hees and P.W.M. Glaudemans, Z. Phys. {\bf 303}
 267 (1981)


\bibitem{Mugambi-70} P.E. Mugambi, Ph.D. Thesis, University of Rochester (1970)

\bibitem{Sarkar-92} S. Sarkar, Ph.D. Thesis, University of Calcutta (1992)

\bibitem{Ormand-97} W.E. Ormand, Phys. Rev. C {\bf 55} 2407 (1997)

\end{thebibliography}
\end{document}